\documentclass[prb,twocolumn,superscriptaddress]{revtex4-1}
\usepackage{amsfonts,amsmath,graphicx,amssymb,bm}
\usepackage[pdftex,plainpages=false,colorlinks=true,linkcolor=blue, citecolor=blue, urlcolor=blue]{hyperref}
\pdfoutput=1

\usepackage[french]{babel}  
\usepackage{boxedminipage}  
\usepackage{float}      
\usepackage[ansinew]{inputenc}  
\usepackage{epsfig}
\usepackage{subfigure}
\usepackage{graphicx}
\usepackage{amsmath}
\usepackage{amssymb}

\renewcommand{\d}[2]{\frac{#1}{#2}}
\newcommand{\pd}{\partial}
\DeclareMathOperator\erf{erf}

\begin{document}

\hyphenation{Brillouin}

\title{Transport functions for hypercubic and Bethe lattices}

\author{Louis-Fran\c cois Arsenault}
\altaffiliation{Present address: Department of Physics, Columbia University, New York, New York 10027, USA}
\author{A.-M. S. Tremblay$^{2,}$}
\affiliation{D\'{e}partement de Physique and RQMP, Universit\'{e} de Sherbrooke, Sherbrooke,
QC J1K 2R1, Canada\\
$^{2}$Canadian Institute for Advanced Research, Toronto, Ontario, Canada.\\}

\date{\today}

\begin{abstract}

In calculations of transport quantities, such as the electrical conductivity, thermal conductivity, Seebeck, Peltier, Nernst, Ettingshausen, Righi-Leduc, or Hall coefficients, sums over the Brillouin zone of wave-vector derivatives of the dispersion relation commonly appear. When the self-energy depends only on frequency, as in single-site dynamical mean-field theory, it is advantageous to perform these sums once and for all. We show here that in the case of a hypercubic lattice in d dimensions, the sums needed for any of the transport coefficients can be expressed as integrals over powers of the energy weighted by the energy-dependent non-interacting density of states. It is also shown that our exact expressions for the transport functions can be obtained from differential equations that follow from sum rules. By substituting the Bethe lattice density of states, one can obtain the previously unknown transport function for the electrical or thermal Hall coefficients and for the Nernst coefficient of the Bethe lattice.
\end{abstract}


\maketitle

\section{Introduction\label{sec:intro}}

The calculation of transport properties is a challenge in the presence of strong correlations since the Boltzmann equation approach is inappropriate in that case. One must fall back on many-body methods. A considerable simplification occurs in the one-band case when the momentum dependence of the self-energy can be neglected since vertex corrections then disappear and the one-particle Green's function suffices for the calculation.

For three-dimensional materials, Dynamical mean-field theory~\cite{Georges:1992,Jarrell:1992,Georges:1996} (DMFT) provides a framework to obtain the needed accurate expressions for the single-particle Green's function with a momentum-independent self-energy. In that case, the calculation of transport coefficients is simplified if one can compute functions $\Phi$ defined by sums over $k$ of the form $\Phi(\varepsilon)=\sum_k F(k)\delta(\varepsilon-\varepsilon_k)$ where $F(k)$ contains derivatives of the dispersion relation $\varepsilon_k$. The calculation of this type of functions, $\Phi(\varepsilon)$, that we will call transport functions, may be difficult even if no interaction effect enters. One can rely on numerical calculations but since DMFT is exact in infinite dimension,~\cite{Metzner:1989,Georges:1992,Jarrell:1992,Georges:1996} it has been customary to rely on results valid in the large dimension limit~\cite{Jarrell:1993,Lange:Kotliar:1998,MerinoMcKenzieTransport:2000,LaadHall:2001}.

We show in this paper that for an hypercubic lattice in any dimension, it is possible to reduce the calculation of the transport functions to integrals of powers of the energy weighted by the non-interacting density of states. In addition, we show that the Hall transport function also follows from a differential equation obtained from a sum rule, in analogy with the case of the conductivity discussed by Chattopadhyay and coworkers.~\cite{Chattopadhyay:Millis:1998} The same transport function occurs in the calculation of the thermal Hall conductivity (Righi-Leduc) and of the Nernst coefficient.

The differential equation approach also allows us to obtain an expression for the Hall transport function that is valid on a Bethe lattice (Cayley tree) in the large coordination limit. As in the case of the conductivity,~\cite{Chattopadhyay:Millis:1998,Chung:Freericks:1998} we show that it is this Hall transport function that should be used, not the one traditionally obtained by substituting the Bethe lattice density of states in the infinite dimensional result.~\cite{Lange:Kotliar:1999} This is important since the DMFT self-consistency relation becomes very simple on the Bethe lattice, making this density of states a very popular choice, even in recent papers.~\cite{Arsenault:2008,MatsuoOkamotoThermopower:2011,HamerlaBethe:2013,PetersKondoBethe:2013,Georges:2013}

In the next section, we define the transport functions. The following section contains the results. A few numerical examples appear before the conclusion. An appendix contains the details of the calculation on the hypercubic lattice.

\section{Definitions of transport functions}
The transport coefficients in the presence of a magnetic field can be defined as in Ziman\cite{Ziman:book}
\begin{eqnarray}
\mathbf{J}_{e} &=&\mathbf{L}_{EE}\left( \mathbf{H}\right) \mathbf{\cdot E+L}%
_{ET}\left( \mathbf{H}\right) \mathbf{\cdot \nabla T}
\label{ElectricalCurrent} \\
\mathbf{J}_{Q} &=&\mathbf{L}_{TE}\left( \mathbf{H}\right) \mathbf{\cdot E+L}%
_{TT}\left( \mathbf{H}\right) \mathbf{\cdot \nabla T}  \label{ThermalCurrent}
\end{eqnarray}
where $\mathbf{J}_{e}$ and $\mathbf{J}_{Q}$ are electrical and heat currents, while $\mathbf{E}$ and $\mathbf{H}$ are electric and magnetic fields respectively and $\mathbf{\nabla T}$ is the temperature gradient. The transport coefficients are matrices. Onsager's reciprocity relations state that~\cite{Ziman:book}
\begin{eqnarray}
\mathbf{L}_{EE}\left( \mathbf{H}\right)  &=&\mathbf{L}_{EE}^{T}\left( -%
\mathbf{H}\right)  \\
\mathbf{L}_{TE}\left( \mathbf{H}\right)  &=&-T\mathbf{L}_{ET}^{T}\left( -%
\mathbf{H}\right)  \\
\mathbf{L}_{TT}\left( \mathbf{H}\right)  &=&\mathbf{L}_{TT}^{T}\left( -%
\mathbf{H}\right)
\end{eqnarray}
where the superscript $T$ indicates the transpose of the matrix.

Let us assume that the magnetic field is applied in the $z$ direction while the currents and the electric field and thermal gradient are in the $xy$ plane. We give below expressions for obtaining the transport coefficients, such as conductivity, in the DC limit, but the same transport functions appear in the corresponding AC expressions. These expressions do not contain vertex corrections. This is justified for one-band models in the infinite dimensional limit~\cite{KhuranaVertex:1990} and in general for wave vector independent self-energies on lattices with inversion symmetry. The expressions are valid to linear order in $H$ (magnetic field) only. We work in units where $\hbar = 1$ and lattice spacing is also unity. We assume an isotropic system for simplicity but the generalization is obvious.

We normalize the spectral weight as follows, $\int A(k,\omega)d\omega=1$. For a local self-energy, the single-particle spectral weight depends on wave vector only through the single-particle dispersion relation, i.e. $A(k,\omega) = A(\varepsilon_k,\omega)$ so that the DC conductivity can be written as~\cite{Jarrell:1993}
\begin{eqnarray}\label{conduc_1}
    \sigma_{xx} &\equiv& L_{EE}^{xx} \\
    &=&\pi e^2\sum_{\sigma}\int d\varepsilon\Phi_{xx}(\varepsilon)\int d\omega \left(-\d{\pd f(\omega)}{\pd \omega}\right)A_{\sigma}^2(\omega,\varepsilon),\nonumber
\end{eqnarray}
where the conductivity transport function is defined by
\begin{equation}\label{Phi_1}
    \Phi_{xx}(\varepsilon) = \sum_k\left(\d{\pd\varepsilon_k}{\pd k_x}\right)^2\delta(\varepsilon-\varepsilon_k).
\end{equation}
The DC, low field, antisymmetric part of the Hall conductivity is given by~\cite{VorugantiJohnHall:1992}
\begin{eqnarray}\label{conduc_Hall_1}
    \sigma_{xy} &\equiv& L_{EE}^{xy} \\
    &=& \d{\pi^2|e|^3}{3}H\sum_{\sigma}\int d\varepsilon\Phi_{xy}(\varepsilon)\int d\omega \left(-\d{\pd f(\omega)}{\pd \omega}\right)A_{\sigma}^3(\omega,\varepsilon),\nonumber
\end{eqnarray}
where $H$ is the magnetic field and the Hall transport function $\Phi_{xy}(\varepsilon)$ is defined by
\begin{equation}\label{Phi_2}
    \Phi_{xy}(\varepsilon) = \sum_k\left[ \d{v_x^2}{M_{yy}} + \d{v_y^2}{M_{xx}} - 2\d{v_xv_y}{M_{xy}}\right]\delta(\varepsilon-\varepsilon_k),
\end{equation}
with $v_{\alpha} = \d{\pd\varepsilon_k}{\pd k_{\alpha}}$ and $M_{\alpha\beta}^{-1} = \d{\pd^2\varepsilon_k}{\pd k_{\alpha}\pd k_{\beta}}$. The above two transport functions, $\Phi_{xx}(\varepsilon)$ and $\Phi_{xy}(\varepsilon)$, suffice to define all other transport coefficients, as we verify below.\\

\begin{widetext}
The diagonal component of the thermoelectric tensor is~\cite{Palsson:Kotliar:1998}
\begin{eqnarray}\label{thermoelec_11}
    \alpha_{xx} \equiv L_{ET}^{xx}
    =-\pi |e|\sum_{\sigma}\int d\varepsilon\Phi_{xx}(\varepsilon)\int d\omega \left(-\d{\pd f(\omega)}{\pd \omega}\right)\left(\frac{\omega}{T}\right)A_{\sigma}^2(\omega,\varepsilon).
\end{eqnarray}
The off-diagonal component on the other hand is given according to Ref.~[\onlinecite{Xu:2013}] by
\begin{eqnarray}\label{thermoelec_12}
    \alpha_{xy} \equiv L_{ET}^{xy}
    = \d{\pi^2|e|^3}{3}H\sum_{\sigma}\int d\varepsilon\Phi_{xy}(\varepsilon)\int d\omega \left(-\d{\pd f(\omega)}{\pd \omega}\right)\left(\frac{\omega}{T}\right)A_{\sigma}^3(\omega,\varepsilon).
\end{eqnarray}

The diagonal component of the response for the thermal transport is
\begin{eqnarray}\label{thermal_11}
    \beta_{xx} \equiv L_{TT}^{xx}
    =-\pi T\sum_{\sigma}\int d\varepsilon\Phi_{xx}(\varepsilon)\int d\omega \left(-\d{\pd f(\omega)}{\pd \omega}\right)\left(\frac{\omega}{T}\right)^2 A_{\sigma}^2(\omega,\varepsilon).
\end{eqnarray}
The vanishing of the vertex corrections for both $\beta_{xx}$ and $\alpha_{xx}$ for momentum-independent self-energies were proven in Ref.~[\onlinecite{Paul:2003}].

The off-diagonal component of the thermal transport can be inferred from the fact that that in the Boltzmann limit for impurity scattering, there is a Wiedemann-Franz law that holds,~\cite{Ziman:book} relating Hall conductivity to transverse thermal transport $\mathbf{L}_{TT}=-\mathit{L_0}T\mathbf{L}_{EE}$ through the Lorentz number $\mathit{L_0}=\pi^2k_B^2/(3e^2)$. From this requirement, we find
\begin{equation}\label{thermal_12}
    \beta_{xy} \equiv L_{TT}^{xy} = -\d{\pi^2|e|T}{3}H \sum_{\sigma}\int d\varepsilon\Phi_{xy}(\varepsilon)\int d\omega \left(-\d{\pd f(\omega)}{\pd \omega}\right)\left(\frac{\omega}{T}\right)^2 A_{\sigma}^3(\omega,\varepsilon).
\end{equation}
\end{widetext}

Finally, the so called stress tensor is given by
\begin{equation}\label{stress_1}
    \tau_{xx} = e^2\sum_{\sigma}\int d\varepsilon\widetilde{\Phi}_{xx}(\varepsilon)\int d\omega f(\omega)A_{\sigma}(\omega,\varepsilon),
\end{equation}
where
\begin{equation}\label{Phi_3}
    \widetilde{\Phi}_{xx}(\varepsilon) = \sum_k\d{\pd^2\varepsilon_k}{\pd k_x^2}\delta(\varepsilon-\varepsilon_k).
\end{equation}
while the off-diagonal stress tensor is defined by
\begin{equation}\label{stress_2}
    \tau_{xy} = \d{He}{2}\sum_{\sigma}\int d\varepsilon\widetilde{\Phi}_{xy}(\varepsilon)\int d\omega f(\omega)A_{\sigma}(\omega,\varepsilon),
\end{equation}
with
\begin{equation}\label{Phi_4}
\widetilde{\Phi}_{xy}(\varepsilon) = 2\sum_k\text{det}\left[\d{\pd^2\varepsilon_k }{\pd k_{\nu} \pd k_{\mu}}\right]\delta (\varepsilon-\varepsilon_k).
\end{equation}

Our primary goal in this paper is to calculate the transport functions $\Phi_{xx}$ and $\Phi_{xy}$. $\widetilde{\Phi}_{xx}$ and $\widetilde{\Phi}_{xy}$ come in sum rules that are important as intermediate steps. We refer to $\Phi_{xx}$ as the conductivity transport function and to $\Phi_{xy}$ as the Hall transport function, but these same transport functions are the only ones that appear in the calculation of any of the transport quantities, as is clear from above. \\

All the usual DC transport coefficients can be obtained from the above expressions and Onsager's relations. For example, following Ziman~\cite{Ziman:book} we define the resistivity tensor
\begin{equation}
\mathbf{\rho }=\mathbf{L}_{EE}^{-1}=\frac{1}{\sigma _{xx}\sigma _{yy}-\sigma
_{xy}\sigma _{yx}}\left(
\begin{array}{cc}
\sigma _{yy} & -\sigma _{xy} \\
-\sigma _{yx} & \sigma _{xx}%
\end{array}%
\right)
\end{equation}
the thermopower tensor
\begin{widetext}
\begin{equation}
\mathbf{Q}=-\mathbf{L}_{EE}^{-1}\mathbf{L}_{ET}=\frac{-1}{\sigma _{xx}\sigma
_{yy}-\sigma _{xy}\sigma _{yx}}\left(
\begin{array}{cc}
\sigma _{yy}\alpha _{xx}-\sigma _{xy}\alpha _{yx} & \sigma _{yy}\alpha
_{xy}-\sigma _{xy}\alpha _{yy} \\
-\sigma _{yx}\alpha _{xx}+\sigma _{xx}\alpha _{yx} & -\sigma _{yx}\alpha
_{xy}+\sigma _{xx}\alpha _{yy}%
\end{array}%
\right)
\end{equation}
\end{widetext}
the Peltier tensor,
\begin{equation}
\mathbf{\Pi }\left( \mathbf{H}\right) =T\mathbf{Q}^{T}\left( -\mathbf{H}%
\right)
\end{equation}%
and, without explicitly doing the matrix multiplications to save space, the thermal conductivity tensor,
\begin{equation}
\mathbf{\kappa =-}\left( \mathbf{L}_{TT}-\mathbf{L}_{TE}\mathbf{L}_{EE}^{-1}%
\mathbf{L}_{ET}\right).
\end{equation}
Some of the better known transport coefficients when the longitudinal current is in the $x$ direction include the Hall resistance $R_H=-\rho_{xy}$, the Seebeck coefficient, or thermopower $Q_{xx}$, the Nernst coefficient, $-Q_{xy}$, and the Righi-Leduc coefficient, $\kappa_{xy}/\kappa_{xx}$.

\section{Transport functions for the Hypercubic lattice in d dimensions}

We show by explicit calculation in Appendix A that for the hypercubic lattice with nearest-neighbor hopping, the transport functions can be expressed in terms of integrals over powers of energy weighted by the non-interacting density of states. The results are:

\begin{equation}\label{Phi_1_2}
    \Phi_{xx}(\varepsilon) = -\d{1}{\text{d}}\int_{-\infty}^{\varepsilon}zN_0(z)dz,
\end{equation}

\begin{equation}\label{Phi_2_1}
    \widetilde{\Phi}_{xx}(\varepsilon) = \d{d\Phi_{xx}(\varepsilon)}{d\varepsilon} = -\d{1}{\text{d}}\varepsilon N_0(\varepsilon),
\end{equation}

\begin{equation}\label{phi_3}
\begin{split}
    \Phi_{xy}(\varepsilon) &= \d{2\varepsilon}{(\text{d}-1)}\Phi_{xx}(\varepsilon)\\ &+ \d{4}{\text{d}(\text{d}-1)}\int_{-\infty}^{\varepsilon}z^2N_0(z)dz\\ &- 2\d{(2t)^2}{(\text{d}-1)}\int_{-\infty}^{\varepsilon}N_0(z)dz,
\end{split}
\end{equation}
where $N_0(z)$ is the non-interacting density of states and d, the dimension.\\
\\
In two dimensions,~\cite{Economou:1983} the non-interacting density of states is given by a complete elliptic integral of the first kind. An analytical expression in terms of complete elliptic integrals of the first and second kind exists for $\Phi_{xx}(\varepsilon)$ Eq.~\eqref{Phi_1_2}. In three dimensions,~\cite{Economou:1983} integrals over elliptic integrals times polynomials of their arguments need to be done numerically.\\
\\
The first application of Eq.~\eqref{conduc_Hall_1} in the context of DMFT can probably be traced back to Ref.~[\onlinecite{Pruschke:anomalous:1995}]. They also considered the hypercubic lattice, but they limited their treatment from the beginning to the special case $\text{d}\rightarrow\infty$ and therefore only obtained the $\text{d}\rightarrow\infty$ limit of Eq.~\eqref{phi_3}.\\
\\

In the rest of this section, we show that we recover the results obtained previously in infinite dimension and also from sum rules for $\Phi_{xx}$. We generalize the latter approach to obtain $\Phi_{xy}$ from a sum rule.

\subsection{The case of infinite dimension}
Here we show that in the limit of infinite dimension we recover the results of Lange and Kotliar \cite{Lange:Kotliar:1999}. When $\text{d} \rightarrow \infty$, one needs to scale hopping as $t\rightarrow \d{t}{\sqrt{2\text{d}}}$ to obtain a finite density of states. Following Ref.~[\onlinecite{Lange:Kotliar:1999}], we also define $t = 1/2$. Thus, the density of states can be written as~\cite{MullerCorrelated:1989} $N_0(z) = \sqrt{\d{2}{\pi}}\text{e}^{-2z^2}$ while $t^2 = \d{1}{8\text{d}}$. For $\Phi_{xx}(\varepsilon)$, the integrand is $z\text{e}^{-2z^2}$ so that as in Refs.~[\onlinecite{Jarrell:1993}] and [\onlinecite{Lange:Kotliar:1999}]
\begin{equation}\label{phi_1_infd}
    \Phi_{xx}(\varepsilon) = \d{1}{4\text{d}}N_0(\varepsilon).
\end{equation}

For $\Phi_{xy}(\varepsilon)$ note that $\int_0^u z^2\text{e}^{-z^2}dz = \d{1}{4}\sqrt{\pi} \erf(u) - \d{1}{2}u\text{e}^{-u^2}$ where $\erf(u)$ is the error function $\int_0^u \text{e}^{-z^2}dz = \d{1}{2}\sqrt{\pi} \erf(u)$. Since $\erf(-\infty) = -1$ and $\text{d}-1 = \text{d}$ as $\text{d} \rightarrow \infty$, we obtain
\begin{equation}\label{phi_3_infd}
    \Phi_{xy}(\varepsilon) = -\d{1}{2\text{d}^2}\varepsilon N_0(\varepsilon).
\end{equation}
Allowing for a factor of $2$ difference in the definitions of $\Phi_{xy}(\varepsilon)$, again we agree with Ref.~[\onlinecite{Lange:Kotliar:1999}].\\
\\

\subsection{Conductivity transport function from sum rule}
Chattopadhyay \emph{et al}. \cite{Chattopadhyay:Millis:1998} argue that the f-sum rule, where $\widetilde{\Phi}_{xx}$ appears, can be used to obtain a differential equation for $\Phi_{xx}$. Indeed, for a tight-binding model, the f-sum rule is given by
\begin{equation}\label{fsum_rule}
    \int_{-\infty}^{\infty}\d{d\omega}{\pi}\text{Re}\{\sigma_{xx}(\omega)\} = \sum_k\d{\pd^2\varepsilon_k}{\pd k_x^2}\langle n_k\rangle.
\end{equation}
But, if only nearest-neighbor hopping is allowed, we have that ${\pd^2\varepsilon_k}/{\pd k_x^2}=-\alpha\varepsilon_k/\text{d}$ where $\alpha$ a constant that depends on the type of lattice. The right hand side of the above equation may thus be rewritten as


\begin{equation*}
    \int d\omega f(\omega)\sum_k\d{\pd^2\varepsilon_k}{\pd k_x^2}A(k,\omega) = -\d{\alpha}{\text{d}}\int d\omega f(\omega)\sum_k\varepsilon_kA(k,\omega),
\end{equation*}
or
\begin{equation}\label{equality_stress_energy}
    \int d\varepsilon\int d\omega f(\omega)\widetilde{\Phi}_{xx}(\varepsilon)A(\varepsilon,\omega) = -\d{\alpha}{\text{d}}\int d\omega f(\omega)N_0(\varepsilon)\varepsilon A(\varepsilon,\omega).
\end{equation}
By imposing that the two are equal, we obtain the result
\begin{equation*}
    \widetilde{\Phi}_{xx}(\varepsilon) = -\d{\alpha}{\text{d}}\varepsilon N_0(\varepsilon)
\end{equation*}
and since we proved in Eq.~\eqref{derive_Phi_1} the general relation $\widetilde{\Phi}_{xx}(\varepsilon) = {d \Phi_{xx}(\varepsilon)}/{d\varepsilon}$, we have that
\begin{equation}\label{diff_form_Phi_1}
    \d{d \Phi_{xx}(\varepsilon)}{d\varepsilon} = -\d{\alpha}{\text{d}}\varepsilon N_0(\varepsilon).
\end{equation}
This result is the same as for the simple cubic lattice Eqs.~\eqref{Phi_1_2} and \eqref{Phi_2_1} since for this lattice $\alpha = 1$. \\
\\
\subsection{Hall transport function from sum rule}
Inspired by these last results we show that one can find the transport function for the Hall conductivity $\Phi_{xy}$ from the sum rule found for the imaginary part of the frequency dependent Hall conductivity by Drew and Coleman\cite{DrewColemanSumRule:1997} and by Lange and Kotliar\cite{Lange:Kotliar:1999} for a tight-binding model:
\begin{equation}\label{Hall_sumrule}
    \int_{-\infty}^{\infty}d\omega\d{\omega\text{Im}\{\sigma_{xy}(\omega)\}}{\pi e^2} = He\sum_k\text{det}\left[\d{\pd^2\varepsilon_k }{\pd k_{\nu}\pd k_{\mu}}\right]\langle n_k\rangle.
\end{equation}
The right-hand side is the off-diagonal stress tensor Eq.~\eqref{stress_2} related to the transport function $\widetilde{\Phi}_{xy}(\varepsilon)$, Eq.~\eqref{Phi_4},
analogous to $\widetilde{\Phi}_{xx}(\varepsilon)$. We will find a relation between $\Phi_{xy}(\varepsilon)$ and this transport function following steps analogous to those used above to relate $\Phi_{xx}(\varepsilon)$ and $\widetilde{\Phi}_{xx}(\varepsilon)$. Starting from the definition of $\Phi_{xy}$ Eq.~\eqref{Phi_2} and following the same steps as in Eq.~\eqref{derive_Phi_1} we first find that
\begin{equation}\label{d_Phi3}
    \d{d\Phi_{xy}(\varepsilon)}{d\varepsilon} = 2\sum_k\text{det}\left[\d{\pd^2\varepsilon_k }{\pd k_{\nu}\pd k_{\mu}}\right]\delta (\varepsilon-\varepsilon_k) = \widetilde{\Phi}_{xy}(\varepsilon).
\end{equation}
Thus, the sum rule for the Hall effect is related to a derivative of $\Phi_{xy}(\varepsilon)$.

Specializing to the hypercubic lattice, the determinant becomes $\text{det}\left[\d{\pd^2\varepsilon_k }{\pd k_{x}\pd k_{y}}\right] = (2t)^2\cos (k_x)\cos (k_y)$ and the right-hand side of the sum rule, using symmetries, can be rewritten using
\begin{equation}\label{det_1}
\begin{split}
    &\sum_k(2t)^2 \cos (k_x)\cos (k_y)\langle n_k\rangle\\ &= \d{1}{\text{d}(\text{d}-1)}\sum_k\varepsilon_k^2\langle n_k\rangle - \d{1}{(\text{d}-1)}\sum_k (2t)^2\cos^2(k_x)\langle n_k\rangle\\
    &= \d{1}{\text{d}(\text{d}-1)}\sum_k\varepsilon_k^2\langle n_k\rangle + \d{1}{(\text{d}-1)}\sum_k(2t)^2\sin^2(k_x)\langle n_k\rangle\\ &- \d{(2t)^2}{(\text{d}-1)}\sum_k\langle n_k\rangle.
\end{split}
\end{equation}
Following the steps of Eqs.~\eqref{fsum_rule}-\eqref{diff_form_Phi_1} we evaluate
\begin{equation}\label{2dphi3}
\begin{split}
    &\sum_k\text{det}\left[\d{\pd^2\varepsilon_k }{\pd k_{\nu}\pd k_{\mu}}\right]\delta (\varepsilon-\varepsilon_k)\\
    &= \d{\varepsilon^2}{\text{d}(\text{d}-1)}\sum_k\delta (\varepsilon-\varepsilon_k)\\
    &+ \d{1}{(\text{d}-1)}\sum_k(2t)^2\sin^2(k_x)\delta (\varepsilon-\varepsilon_k)\\
    &- \d{(2t)^2}{(\text{d}-1)}\sum_k\delta (\varepsilon-\varepsilon_k).
\end{split}
\end{equation}
which allows us to find an expression for the derivative of $\Phi_{xy}$, Eq.~\eqref{d_Phi3}, that follows from the sum rule
\begin{equation}\label{d_Phi3_1}
    \d{1}{2}\d{d\Phi_{xy}(\varepsilon)}{d\varepsilon} = \d{1}{\text{d}(\text{d}-1)}\varepsilon^2N_0(\varepsilon) + \d{1}{(\text{d}-1)}\Phi_{xx}(\varepsilon) - \d{(2t)^2}{(\text{d}-1)}N_0(\varepsilon).
\end{equation}
If we take the derivative of Eq.~\eqref{phi_3} for $\Phi_{xy}$ that was obtained for the hypercubic lattice, and use Eq.~\eqref{Phi_2_1}, we obtain the same answer. The result for $\Phi_{xy}$ may thus be obtained directly or from a differential equation that follows from a sum rule.

Unfortunately, the relation between the derivative of $\Phi_{xy}$ and the sum rule for the Hall conductivity Eq.~\eqref{Hall_sumrule} does not seem to hold for all nearest-neighboor dispersion relations, contrary to the f-sum rule which is always proportional to the kinetic energy per direction.  Nevertheless, while the two expressions we obtained independently for $\Phi_{xy}$ and for its derivative are automatically valid on the hypercubic lattice in any dimension, on the Bethe lattice enforcing both of them will uniquely determine the transport function $\Phi_{xy}$ on that lattice.  \\

\subsection{Interacting transport functions}

All our transport coefficients Eqs.~\eqref{conduc_1}, \eqref{conduc_Hall_1}, \eqref{thermoelec_11}, \eqref{thermoelec_12}, \eqref{thermal_11} and \eqref{thermal_12} can also be written in term of two so-called interacting transport functions $\Phi_{tr}^L(\omega)$ and $\Phi_{tr}^T(\omega)$, where $L$ and $T$ stand for longitudinal and transverse. These are defined by
\begin{equation}\label{tautr}
  \Phi_{tr}^L(\omega) \equiv \int d\varepsilon\Phi_{xx}(\varepsilon)A^2(\omega,\varepsilon)
\end{equation}
and
\begin{equation}\label{tautrH}
  \Phi_{tr}^T(\omega) \equiv \int d\varepsilon\Phi_{xy}(\varepsilon)A^3(\omega,\varepsilon).
\end{equation}
The definition of the longitudinal interacting transport function, which enters the transport coefficients Eqs.~\eqref{conduc_1}, \eqref{thermoelec_11} and \eqref{thermal_11}, leads to a compact expression commonly used in works by Freericks and coworkers (see for example Ref.[\onlinecite{Freericks:Zlatic:2003}]). The general result for the cubic lattice in d dimensions, re-derived in Appendix B, is given by
\begin{equation}\label{tau_tr}
\begin{split}
  \Phi_{tr}^L(\omega) = \d{1}{2\pi^2}\Bigg(& \d{\text{Im}\left\{G_{tr}^L(z)\right\}}{\text{Im}\left\{z^*\right\}} +\d{1}{\text{d}}\\ &- \d{1}{\text{d}}\text{Re}\left\{zG(z)\right\} \Bigg),
\end{split}
\end{equation}
where $z \equiv \omega + \mu - \Sigma (\omega)$ and

\begin{equation}\label{GLtr}
G_{tr}^L(z) \equiv \int d\varepsilon \d{\Phi_{xx}(\varepsilon)}{z-\varepsilon}~~\\
;~~G(z) \equiv \int d\varepsilon \d{N_0(\varepsilon)}{z-\varepsilon}.\\
\end{equation}

We show in Appendix B that while we can also do a similar analysis for $\Phi_{tr}^T(\omega)$ that enters the transverse transport coefficients Eqs.~\eqref{conduc_Hall_1},\eqref{thermoelec_12}\eqref{thermal_12}, the result is not compact. Therefore, its usefulness is not really clear, except that the first term is related to the usual Boltzmann-like term that depends upon the square of the quasi-particle scattering time $\tau_{Qp}^2 = \left(\d{1}{\text{Im}\{\Sigma\}}\right)^2$. The full result is,
\begin{widetext}
\begin{equation}\label{tau_tr_H}
\begin{split}
  \Phi_{tr}^T(\omega) &= \d{1}{\pi^3}\Bigg[ \d{-3}{8\left(\text{Im}\{z\}\right)^2}\text{Im}\left\{G_{tr}^T(z)\right\} + \d{1}{4\text{d}(\text{d}-1)}\text{Im}\left\{zG(z)\right\}
  +\d{1}{4(\text{d}-1)}\text{Im}\left\{ \left( \d{z^2}{\text{d}} -(2t)^2 \right)\d{\pd G(z)}{\pd z}   \right\}\\
  &+\d{3}{8\text{Im}\{z\}}\left( -\d{2}{\text{d}(\text{d}-1)}\text{Re}\{z\} + \d{2}{\text{d}(\text{d}-1)}\text{Re}\left\{z^2G(z)\right\} + \d{2}{\text{d}-1}\text{Re}\left\{G_{tr}^L(z)\right\} - \d{2(2t)^2}{\text{d}-1}\text{Re}\{G(z)\}      \right)\Bigg],
\end{split}
\end{equation}
\end{widetext}
where

\begin{equation}
G_{tr}^T(z) \equiv \int d\varepsilon \d{\Phi_{xy}(\varepsilon)}{z-\varepsilon}.
\end{equation}

\section{Transport functions for the Bethe lattice}
The self-consistency relation in single site dynamical mean field theory, which is exact in infinite dimensions, is very simple on a Bethe lattice (Caley tree) in the infinite coordination limit i.e. a semi-circular density of states. Hence, this is a commonly used lattice. It has a tree structure with no loop and no dispersion relation in $k$ space. Therefore one has to do all calculations in energy space. But how can we define the transport functions on this lattice? The answer is known for the conductivity transport function $\Phi_{xx}$. In this section, after we describe two methods to obtain the answer for $\Phi_{xx}$ and $\widetilde{\Phi}_{xx}$, we generalize the approach for the Hall transport function $\Phi_{xy}$.

A popular way to obtain $\Phi_{xx}$ for the Bethe lattice has been to a) start from the standard formula in k-space and rewrite it in terms of $\varepsilon$ space; b) take for $\Phi_{xx}(\varepsilon)$ its $\text{d} = \infty$ form (Eq.~\eqref{phi_1_infd}); c) replace $N_0(\varepsilon)$ by the Bethe lattice density of state. But it has been pointed out by Chung and Freericks \cite{Chung:Freericks:1998} and  Chattopadhyay and coworker \cite{Chattopadhyay:Millis:1998} that this does not gives a correct answer. Indeed, the f sum-rule is not satisfied because $\widetilde{\Phi}_{xx}$ in that case is not equal to the derivative of $\Phi_{xx}$. It is in that context that Chattopadhyay and coworker \cite{Chattopadhyay:Millis:1998} proposed their differential equation.

The other way that we propose to obtain the correct result relies on the fact that, as follows from Eq.~\eqref{derive_Phi_1}, the exact result that we obtained on the hypercubic lattice Eq.~\eqref{Phi_1_2} satisfies the differential equation Eq.~\eqref{Phi_2_1} for any nearest-neighbor density of states $N_0(\varepsilon)$. Hence substituting the Bethe lattice density of states in Eq.~\eqref{Phi_1_2} for $\Phi_{xx}$ should also give the correct expression. That approach works essentially because the usual Bethe lattice is also defined by nearest-neighbor hopping only.

In summary, the first derivation of the transport function~\cite{Jarrell:1993} $\Phi_{xx}(\varepsilon)$ took the infinite dimension limit first and then replaced the density of states by that of the Bethe lattice, leading to a failure of the f sum-rule. The other procedure, that satisfies the f sum-rule, is to take the exact result for the transport function valid on the hypercubic lattice for any dimension, and then replace the density of states by that of the Bethe lattice at the end, making sure that $\widetilde{\Phi}_{xx}$ is obtained by a derivative of $\Phi_{xx}$, as in Eq.~\eqref{Phi_1_2}. The correct result would have been obtained already in Ref.~[\onlinecite{Jarrell:1993}] if the calculation had been pushed to the end in finite dimension for all transport functions before taking the infinite-dimensional limit.

The derivation of the known result for $\Phi_{xx}(\varepsilon)$ on the Bethe lattice will allow us to obtain the expression for $\Phi_{xy}(\varepsilon)$. Starting from Eq.~\eqref{Phi_1_2} for $\Phi_{xx}(\varepsilon)$ on the hypercubic lattice, we replace the density of states $N_0(\varepsilon)$ by the Bethe lattice density of states $N_0(\varepsilon) = \d{2}{\pi W^2}\Theta (W-|\varepsilon|)\sqrt{W^2-\varepsilon^2}$ where $W$ is half the bandwidth, and we find
\begin{equation}\label{Phi_1_3}
\begin{split}
    \Phi_{xx}(\varepsilon) &= -\d{1}{d}\int_{-W}^{\varepsilon\leq W}\d{2}{\pi W^2}z\sqrt{W^2-z^2}dz\\
    &= \d{1}{3\text{d}}(W^2-\varepsilon^2)N_0(\varepsilon).
\end{split}
\end{equation}
This agrees with Refs.~[\onlinecite{Chattopadhyay:Millis:1998}] and [\onlinecite{Chung:Freericks:1998}].

To compute $\Phi_{xy}(\varepsilon)$ we also need
\begin{equation}\label{intz2N0}
\begin{split}
    \int_{-\infty}^{\varepsilon}z^2N_0(z)dz &= \d{2}{\pi W^2}\int_{-W}^{\varepsilon\leq W}z^2\sqrt{W^2-z^2}dz\\
    &= -\d{\varepsilon (W^2-2\varepsilon^2)}{8}N_0(\varepsilon) + \d{W^2}{8}\\
    &+ \d{W^2}{4\pi}\tan^{-1}\left(\d{\varepsilon}{\sqrt{W^2-\varepsilon^2}}\right)
\end{split}
\end{equation}
and
\begin{equation}\label{intN0}
\begin{split}
    \int_{-\infty}^{\varepsilon}N_0(z)dz &= \d{2}{\pi W^2}\int_{-W}^{\varepsilon\leq W}\sqrt{W^2-z^2}dz\\
    &= \d{1}{2}\varepsilon N_0(\varepsilon) + \d{1}{2}\\
    &+ \d{1}{\pi}\tan^{-1}\left(\d{\varepsilon}{\sqrt{W^2-\varepsilon^2}}\right).
\end{split}
\end{equation}
We can now substitute Eqs.~\eqref{Phi_1_3} to \eqref{intN0} in the hypercubic lattice expression for $\Phi_{xy}(\varepsilon)$, Eq.\eqref{phi_3}. In that equation there is a term with $(2t)^2$ that we need to define on the Bethe lattice. If we take the connectivity $K$ of the large connectivity Bethe lattice to be equal to $2\text{d}$, we have that\cite{Economou:1983} $W=2t\sqrt{2\text{d}}$. We can also relate $W$ and $t$ by the requirement that the function $\Phi_{xy}$ must change sign at $\varepsilon=0$ along with the change between hole-like and electron-like excitations. If $\Phi_{xy}$ has to be zero at $\varepsilon = 0$ it means that the constant terms in Eqs.\eqref{intz2N0} and \eqref{intN0} must cancel each other once inserted in Eq.\eqref{phi_3}. This is the case if $2t = \d{W}{\sqrt{2\text{d}}}$, the same result as the one found above. With this value of $2t$, the terms with $\tan^{-1}\left(\d{\varepsilon}{\sqrt{W^2-\varepsilon^2}}\right)$ also cancel out so that we finally obtain

\begin{equation}\label{Phi_2_2}
    \Phi_{xy}(\varepsilon) = -\d{1}{3\text{d}(\text{d}-1)}\varepsilon (W^2-\varepsilon^2)N_0(\varepsilon).
\end{equation}
This expression also satisfies the differential equation Eq.~\eqref{d_Phi3_1}. It was used in Ref.~[\onlinecite{Arsenault:2008}] where the Bethe lattice density of states was used as an approximation of the 3d simple cubic lattice. In this reference, there was a small error in the evaluation of $\Phi_{xy}(\varepsilon)$ but the particular cancellations present for the Bethe lattice lead to a final answer that is correct.

\section{Numerical examples}
To illustrate how various approximations for the transport functions differ, we consider the Hall coefficient and the thermopower in a doped Mott insulator on a cubic lattice in the Fermi liquid regime, where the answer depends mostly on the transport function.

\subsection{Approximations for the transport functions}
\begin{figure}
\subfigure{\includegraphics[scale=0.5,angle=0]{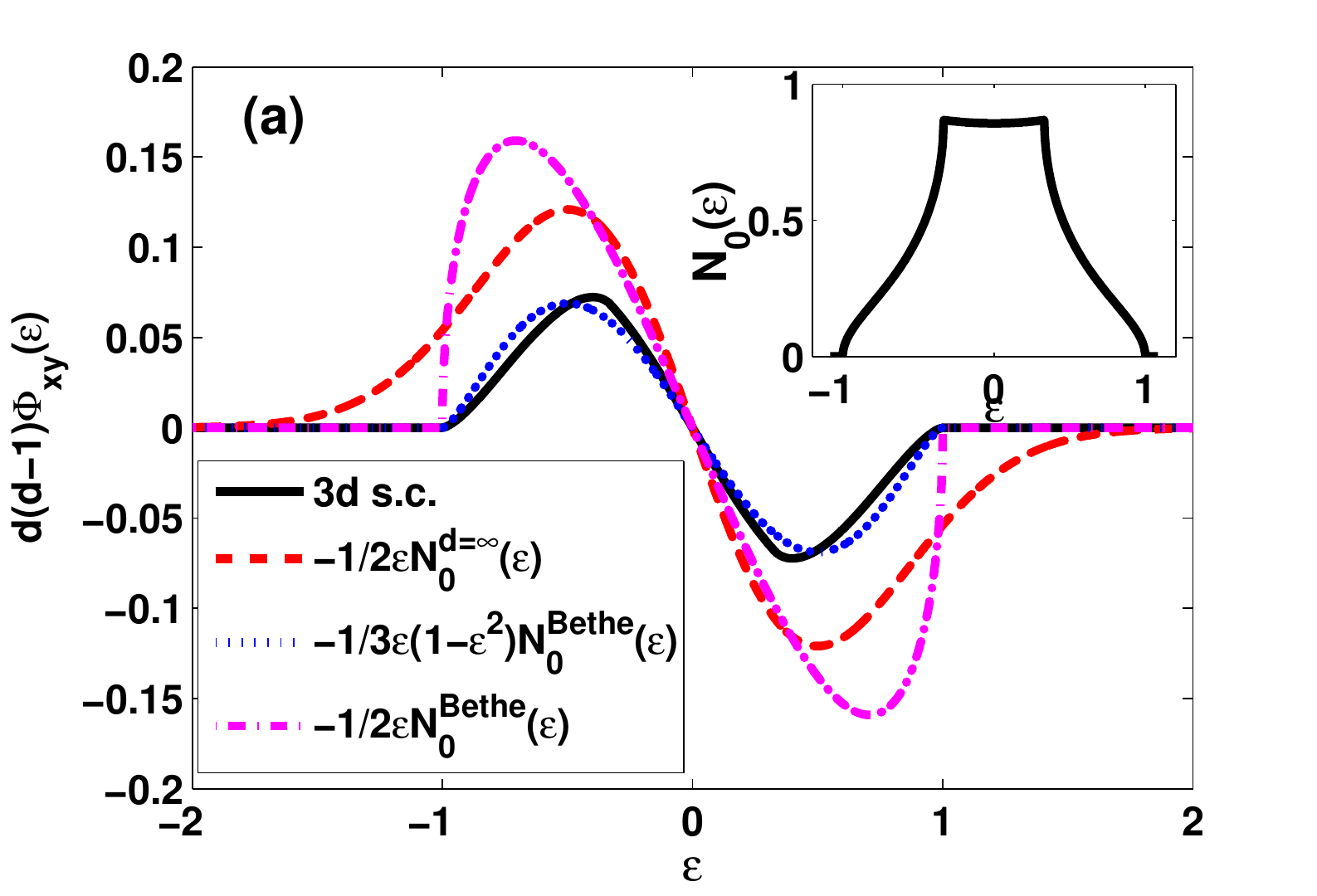}}
\subfigure{\includegraphics[scale=0.5,angle=0]{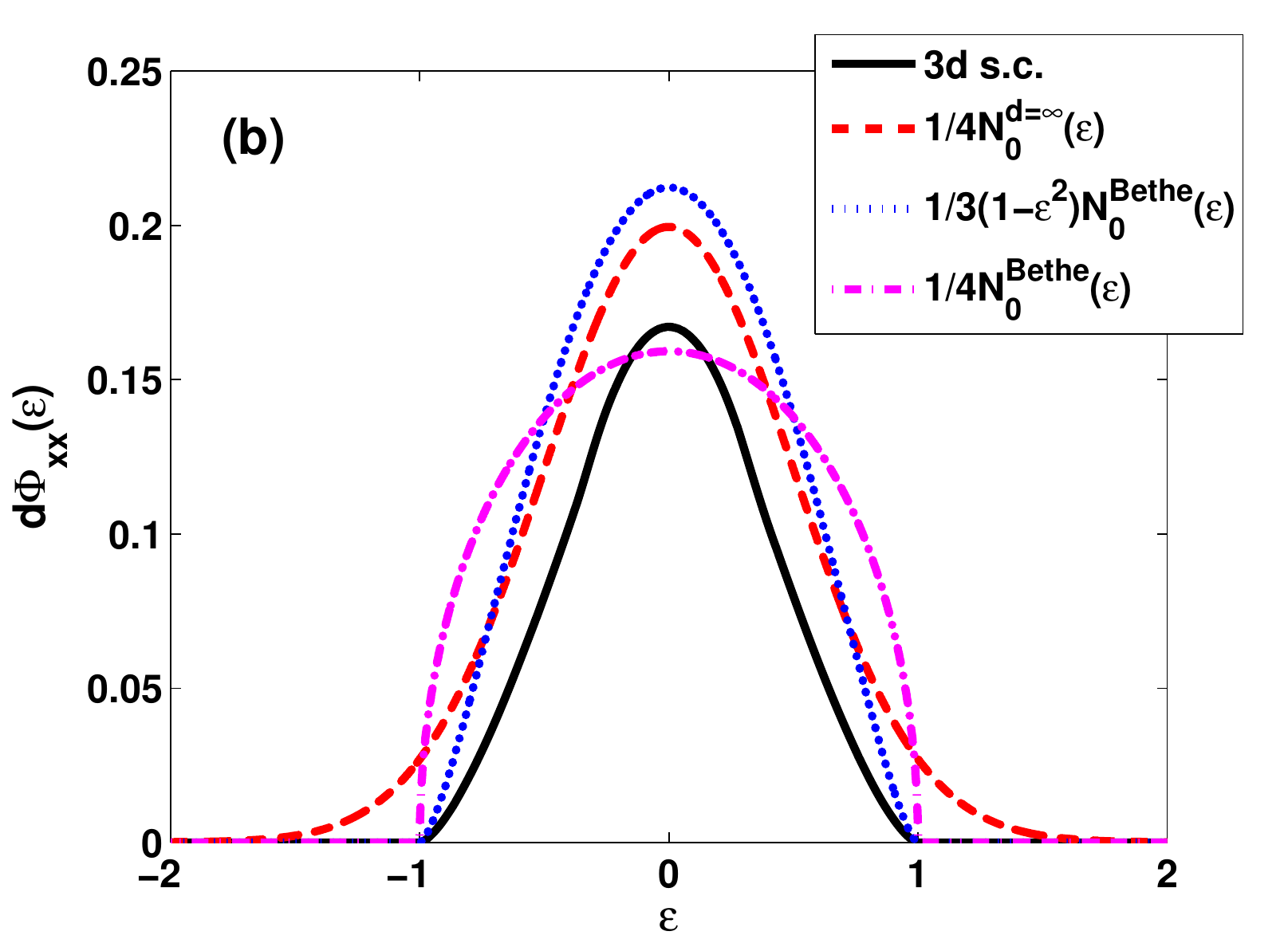}}
\caption{a) $\text{d(d-1)}\Phi_{xy}(\varepsilon)$ for the simple cubic lattic (black solid line), for the infinite-dimensional hypercubic lattice (red dash), for our expression for the Bethe lattice (blue small dash) and for the Bethe lattice density of states substituted in the infinite dimensional result (magnenta short-long dash). b) With the same symbols, the corresponding approximations for $\text{d}\Phi_{xx}(\varepsilon)$. The inset in (a) is the density of states for the simple cubic lattice. }
\label{fig:Phi3}
\end{figure}

First, let us discuss directly various approximations for $\Phi_{xy}$. The density of states of the simple cubic lattice appears in the inset of Fig.~\ref{fig:Phi3}-(a). We show different choices of $\Phi_{xy}$ times d(d-1) as a function of energy $\varepsilon$ normalized such that $W=1$.  $\Phi_{xy}(\varepsilon)$ calculated for the three-dimensional simple cubic lattice, Eq.~\eqref{Phi_1_2}, is represented by the black solid line. The infinite dimensional hypercubic lattice result,  Eq.~\eqref{phi_3_infd}, shown as the red dashed curve is always larger. Recall that in the limit of large dimension d, $\text{d(d-1)}=\text{d}^2$ so it is still correct to call the y axis $\text{d}(\text{d}-1)\Phi_{xy}$. The blue small-dashed curve is the result for large coordination Bethe lattice that we obtained in Eq.~\eqref{Phi_2_2}. Clearly, it is an excellent approximation to the Hall transport function of the simple cubic lattice. Finally, the magenta short-long dashed curve represents the result when one substitutes, as in Ref.~[\onlinecite{Lange:Kotliar:1999}], the Bethe lattice DOS in the $\text{d} = \infty$ result, as was popular for $\Phi_{xx}$ before the work of Refs.~[\onlinecite{Chattopadhyay:Millis:1998}] and [\onlinecite{Chung:Freericks:1998}]. This can be a rather bad approximation.

For completeness, we also show in Fig.~\ref{fig:Phi3}-(b) the different calculations for $\Phi_{xx}$ that were already discussed in the past. Using the Bethe lattice density of states in the expression for the infinite dimensional DOS is not the best choice. The correct choice for the Bethe lattice Eq.~\eqref{Phi_1_3}, shown by the small-dashed blue line, also differs quite substantially from the black solid line for the simple cubic lattice, contrary to the case of the Hall coefficient.

\subsection{Hall coefficient}
To illustrate the effect of the choice of the transport function on the Hall coefficient, we consider a simple, mostly analytic example. Indeed, at very low temperature, DMFT predicts a Fermi-liquid peak at zero frequency. As far as transport is concerned, the derivative of the Fermi-Dirac function guarantees that only the quasi-particle peak is important at low $T$. Thus, one can consider only the Fermi-Liquid self-energy and neglect the incoherent contributions. The self-energy in this regime is given by a Taylor expansion in power of $\omega$. The simplest meaningful (capturing some lifetime effects) case is to stop at second order. In that case, one finds\cite{Lange:Kotliar:1999} $\Sigma (\omega) = (1-1/Z)\omega+\alpha\omega^2 + i\gamma (\omega)$. The constants are $Z = (1-\pd\text{Re}\{\Sigma (\omega)\}/\pd\omega|_{\omega=0})^{-1}$, $\alpha = \d{1}{2}\pd^2\text{Re}\{\Sigma (\omega)\}/\pd\omega^2|_{\omega = 0}$ and the quasiparticle lifetime is $\gamma (\omega) = \tilde{\gamma}(T^2 + (\omega/\pi)^2)$. It should be noted that this form for the self-energy holds only at very low temperature in DMFT calculations even when a symmetric DOS such as that of the Bethe lattice~\cite{Georges:2013} is used. Corrections from $\omega T^2$ and $\omega^3$ terms become important very quickly when temperature is risen~\cite{Georges:2013,Arsenault:thermo:2012,Arsenault:theses}. With this form of self-energy, Lange and Kotliar \cite{Lange:Kotliar:1999} have shown that the Hall constant is
\begin{equation}\label{RH}
    R_H =\d{0.2630}{e} \d{\Phi_{xy} (\tilde{\mu})}{[\Phi_{xx} (\tilde{\mu})]^2},
\end{equation}
where $e$ is the electron charge and $\tilde{\mu} = \mu - \text{Re}\{\Sigma (0)\}$ which, invoking Luttinger's theorem, corresponds to the chemical potential needed to have the same density with the non-interacting density of states. The above result is independent of Fermi liquid parameters and we do not need to solve the full DMFT equations to find $R_H$. The final answer depends only on transport functions.

\begin{figure}
\includegraphics[scale=0.5,angle=0]{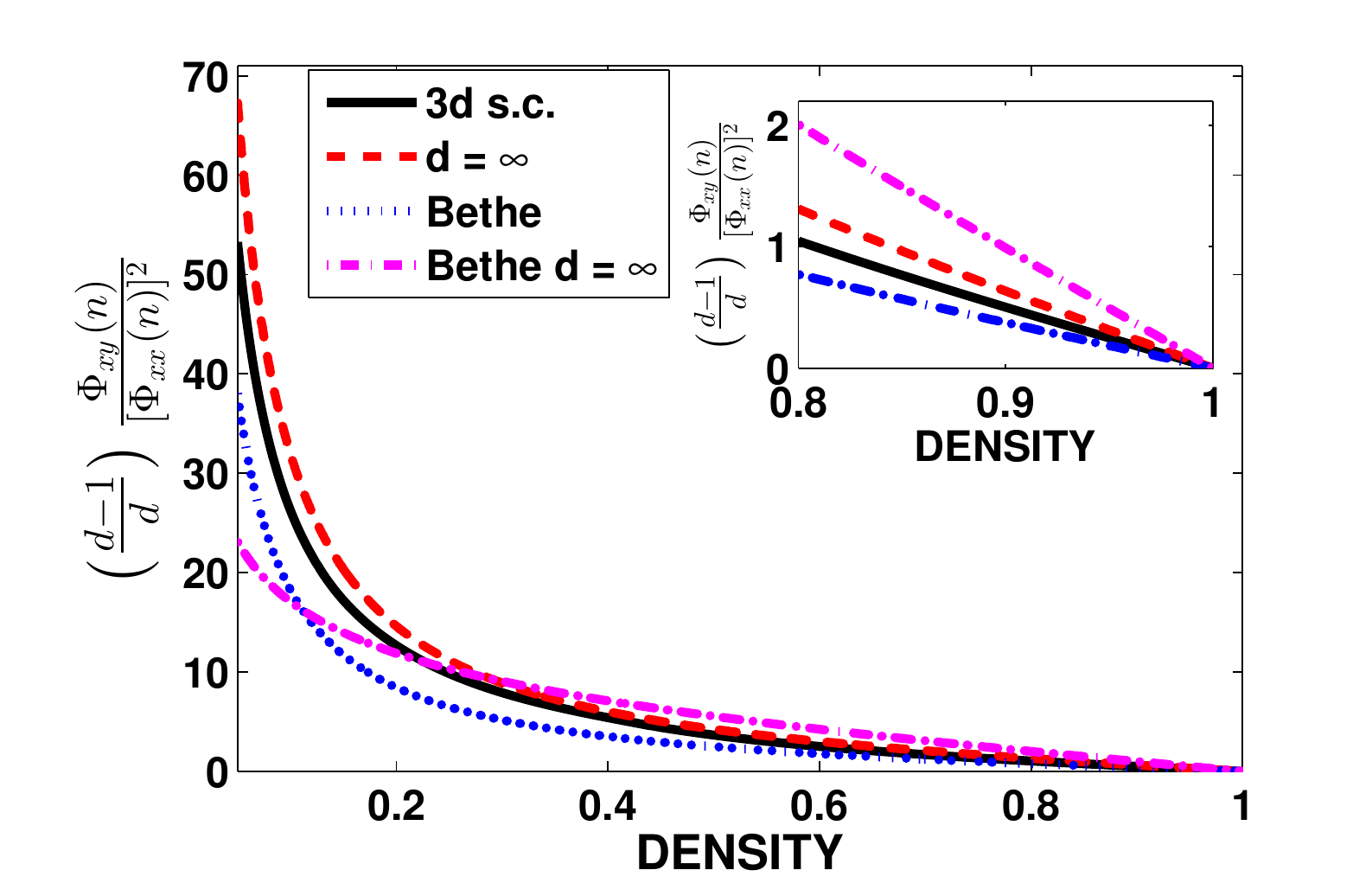}
\caption{Prefactor of the Hall resistance for various transport functions in the Fermi liquid regime of a doped Mott insulator, as a function of density or filling, ($n=1$ being half-filling). Same conventions as in Fig.~\ref{fig:Phi3}: simple cubic lattic (black solid line), the infinite-dimensional hypercubic lattice (red dash), our expression for the Bethe lattice (blue small dash) and the Bethe lattice density of states substituted in the infinite dimensional result (magnenta short-long dash). }
\label{fig:RH}
\end{figure}

Fig.~\ref{fig:RH} displays the various results. $R_H$ being antisymmetric with respect to half-filling, we only show the results for $n\leq 1$. We observe a cancellation of errors in the ratio of transport functions that leads to only relatively small discrepancies between the 3d cubic lattice result, obtained from Eqs.~\eqref{Phi_1_2} and \eqref{phi_3}, (black solid line) and two approximations, namely the infinite dimensional result (red dash) obtained from Eqs.~\eqref{phi_1_infd},\eqref{phi_3_infd} and the sum-rule satisfying expressions for the Bethe lattice (blue short dash), obtained from Eqs.~\eqref{Phi_1_3} and \eqref{Phi_2_2}. As could be seen already in Fig.~\ref{fig:Phi3}, the largest discrepancies occur near the band edges. In this low temperature limit, the Hall coefficient recovers its non-interacting value.~\cite{Lange:Kotliar:1999} Near half-filling $n=1$, the curves are mostly linear as expected from $R_H~\propto 1/n$ but with different slopes (see Inset of Fig.~\ref{fig:RH}). The replacement of the density of states in the infinite dimensional result by the Bethe lattice density result (magenta short-long dash) is an approximation that is unphysical at low density and has the largest discrepancy in its slope near $n=1$. \\

\subsection{Thermopower}
As a last example, consider thermopower. For the quadratic Fermi liquid self-energy, Palsson and Kotliar\cite{Palsson:Kotliar:1998} obtained for the Seebeck coefficient,
\begin{equation}\label{S_FL}
  S = -\d{k_B}{|e|}\d{E_2}{E_0}\d{k_BT}{Z}\d{\Phi_{xx}^{'}(\tilde{\mu})}{\Phi_{xx}(\tilde{\mu})},
\end{equation}
where $E_2$ and $E_0$ are universal positive constants. Contrary to the Hall coefficient, the thermopower depends on a Fermi liquid parameter, the quasiparticle weight $Z$. Since we assume a doped Mott insulator, we take (for all density of states) a simple model where $Z$ is proportional to the doping $x = |1-n|$, which gives $Z = 0$ at half-filling. Hence, we can write
\begin{equation}\label{S_FL_prop}
  \d{S}{T} \propto -\d{1}{|1-n|}\d{\Phi_{xx}^{'}(\tilde{\mu}(n))}{\Phi_{xx}(\tilde{\mu}(n))}.
\end{equation}
The comparisons that we will make will thus be only very qualitative since, normally, in DMFT the precise value of $Z$ depends on the density of states that is used and contains some information about particle-hole asymmetry~\cite{Arsenault:IPT:2012}.

For a simple cubic lattice (or any single band with nearest-neighbor hopping), we may take the derivative of $\Phi_{xx}$, Eq.~\eqref{Phi_1_2}, and write
\begin{equation}\label{S_FL_prop1}
  \d{S}{T} \propto \d{1}{Z}\d{\d{1}{\text{d}}\tilde{\mu}N_0(\tilde{\mu})}{\Phi_{xx}(\tilde{\mu})}.
\end{equation}
The density of states $N_0$, the transport function $\Phi_{xx}$ and the quasiparticle weight $Z$ are all positive and thus, as expected for a particle-hole symmetric system, $S$ changes sign at $n=1$ the value for which $\tilde{\mu} = 0$. For a correlated metal, there is no ambiguity as $Z \neq 0$ at half-filling and thus $S=0$. But, for a Mott insulator, at $n=1$, $Z = \tilde{\mu} = 0$. Using l'Hôpital's rule in Eq.~\eqref{S_FL_prop1} then leads to
\begin{equation}\label{lim_S}
  \lim_{\tilde{\mu} \rightarrow 0, n \rightarrow 1}\d{S}{T} = \d{\d{1}{\text{d}}N_0(0)}{\Phi_{xx}(0)Z'_{\tilde{\mu} = 0, n = 1}}.
\end{equation}
Thus, the limiting behavior is determined by the derivative of $Z$ at half-filling since $Z' = \d{dZ}{dn}\d{dn}{d\tilde{\mu}} = 2N_0(\tilde{\mu})\d{dZ}{dn}$. Taking $Z$ exactly equal to $|1-n|$ for definiteness, the left and right derivatives differ at $n=1$, leading to a discontinuous jump of $S$ from negative to positive at half-filling with values
\begin{equation}\label{lim_S1}
  \lim_{\tilde{\mu} \rightarrow 0, n \rightarrow 1}\d{S}{T} = \d{\mp 1}{2\text{d}\Phi_{xx}(0)}.
\end{equation}

\begin{figure}[tpb]
\subfigure{\includegraphics[scale=0.5,angle=0]{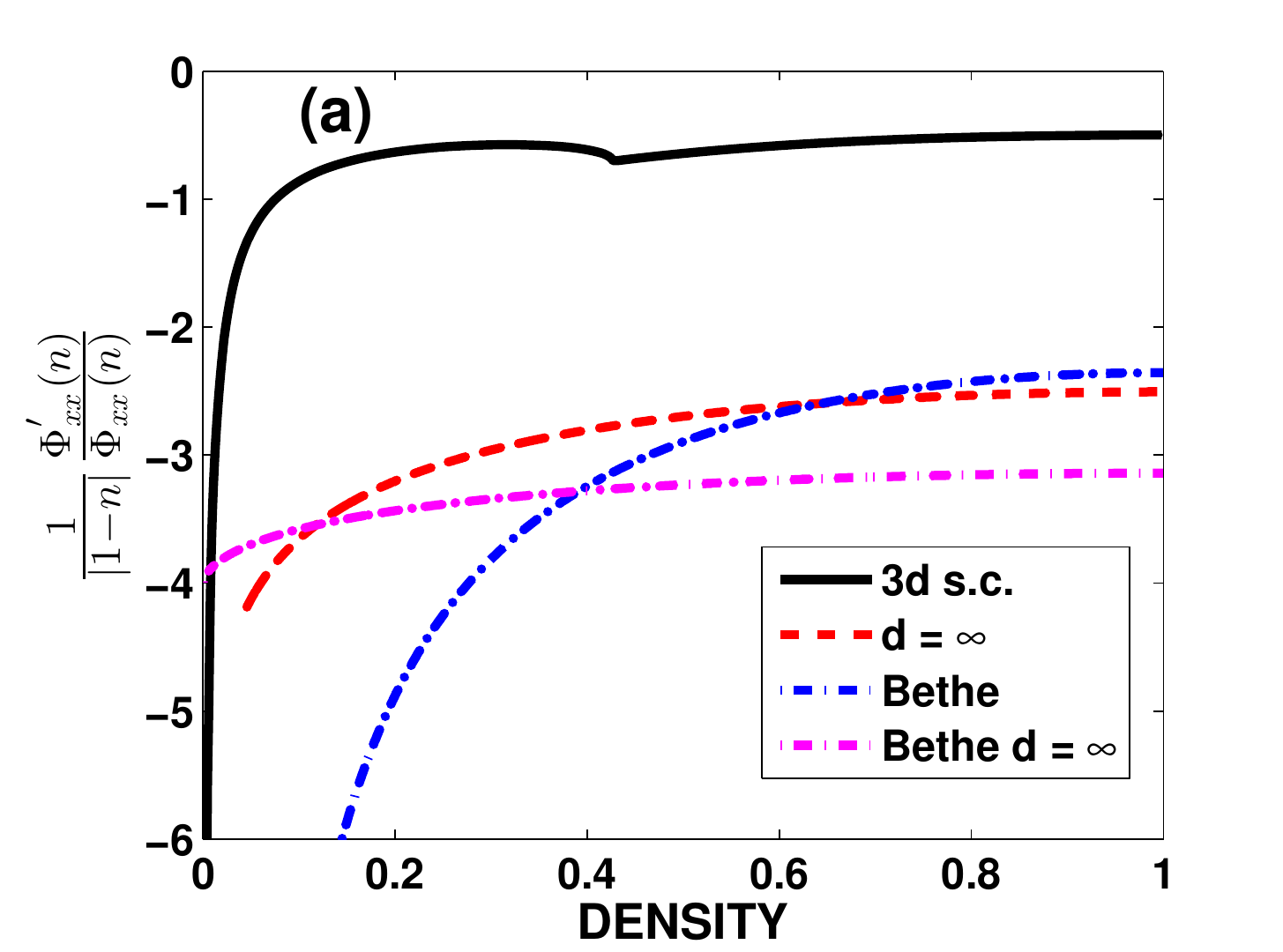}}
\subfigure{\includegraphics[scale=0.5,angle=0]{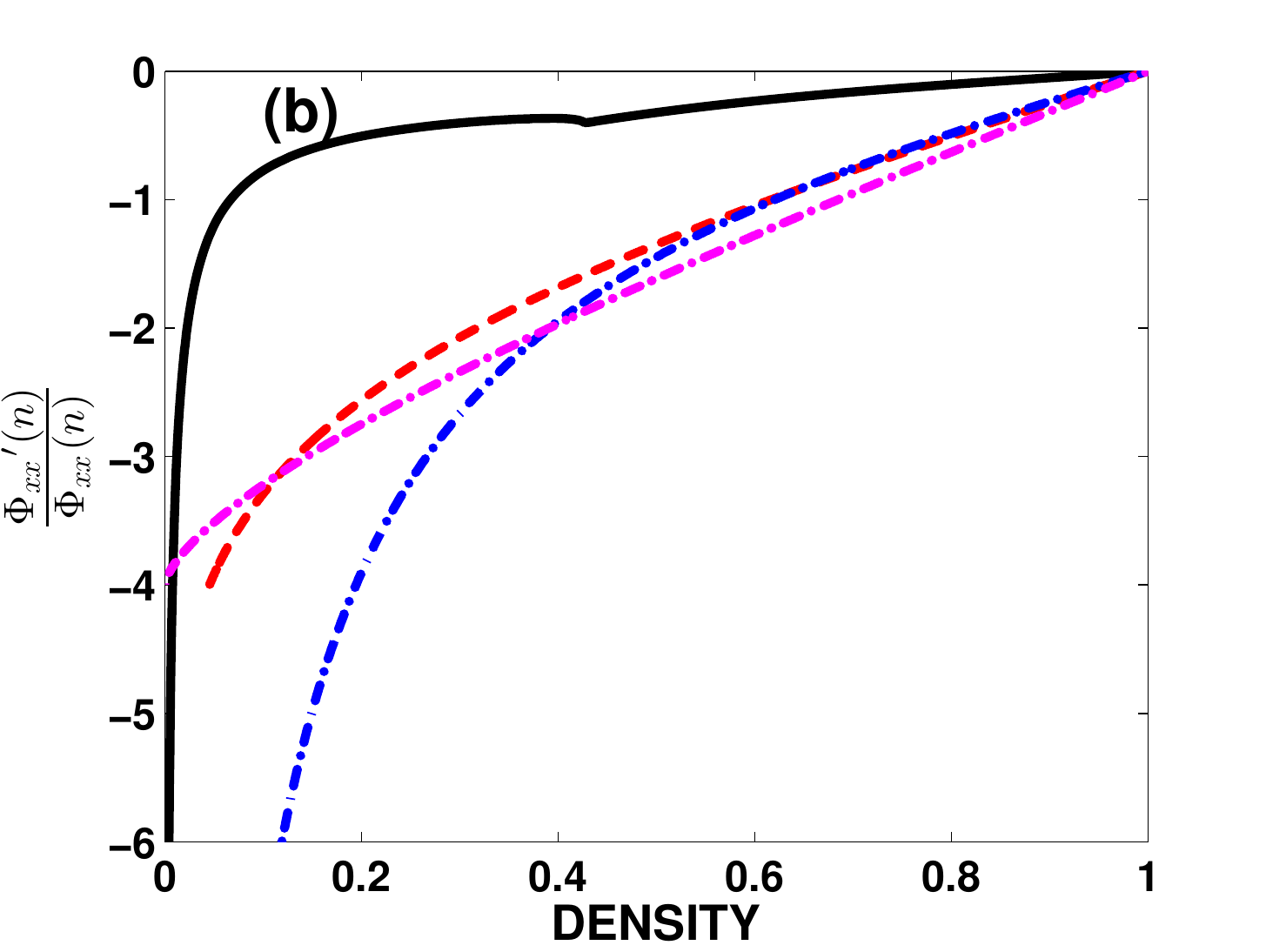}}
\caption{a) Limiting low temperature behavior of the thermopower $\d{S}{T}$ on an arbitrary scale when the system is a Mott insulator at half-filling. Same conventions as in Fig.~\ref{fig:Phi3}: simple cubic lattic (black solid line), the infinite-dimensional hypercubic lattice (red dash), our expression for the Bethe lattice (blue small dash) and the Bethe lattice density of states substituted in the infinite dimensional result (magnenta short-long dash). b) Same as a) but with the prefactor $1/|1-n|$ removed.}
\label{fig:S_T}
\end{figure}

Fig.~\ref{fig:S_T}-(a) displays the results for the thermopower Eq.~\eqref{S_FL_prop1} using the different definitions of the transport function $\Phi_{xx}$. $S$ being antisymmetric with respect to half-filling, we only show the results for $n\leq 1$. In Fig.~\ref{fig:S_T}-(b), we show only $\Phi_{xx}'/\Phi_{xx}$. The results for the cubic lattice are far from the results for all the other transport functions. Like for the Hall coefficient, the case (magenta short-long dash) where the Bethe lattice density of states is used in the infinite dimensional result Eq.~\eqref{phi_1_infd} gives the worse results, especially at low density where it does not reproduce the large thermopower. Hence, the inadequacy of this approximation for the conductivity noted in Refs.~[\onlinecite{Chattopadhyay:Millis:1998}] and [\onlinecite{Chung:Freericks:1998}] carries over to the thermopower and calls for great caution in the interpretation of calculations based on this approximation.~\cite{MatsuoOkamotoThermopower:2011}

\section{Conclusion}
We have found simple expressions for transport functions on hypercubic lattices in terms of integrals of power laws of the energy weighted by the non-interacting density of states, Eqs.~\eqref{Phi_1_2} and \eqref{phi_3}. In particular, we recover  the known results for infinite dimension, Eqs.~\eqref{phi_1_infd} and \eqref{phi_3_infd}. On hypercubic lattices, we have also shown that it is possible to use sum rules to obtain differential equations not only for the conductivity transport function,~\cite{Chattopadhyay:Millis:1998} Eq.~\eqref{Phi_2_1}, but also for the Hall conductivity transport function, Eq.~\eqref{d_Phi3_1}.  This has allowed us to generalize the approach of A. Chattopadhyay \emph{et al}. \cite{Chattopadhyay:Millis:1998} and W. Chung \emph{et al}. \cite{Chung:Freericks:1998} to calculate the transport function for the Hall conductivity on the Bethe lattice, Eq.~\eqref{Phi_2_2}. Numerical calculations for doped Mott insulators in three dimensions in the Fermi liquid regime show that the latter transport function, Eq.~\eqref{Phi_2_2}, gives a good approximation for the result on the simple cubic lattice. These calculations also show that substituting the Bethe lattice density of states in the infinite dimensional result, Eq.~\eqref{phi_3_infd}, should be avoided. The two transport functions that we have calculated are the only ones that are necessary to obtain all the transport coefficients.

\acknowledgments
We acknowledge J. Freericks and G. Kotliar for discussions. This work was supported by the Natural Sciences and Engineering Research Council of Canada (NSERC) (L.-F.A. and A.-M.S.T), and by the Tier I Canada Research Chair Program (A.-M.S.T.). Simulations were performed on computers provided by CFI, MELS, Calcul Qu\'ebec and Compute Canada.

\appendix
\section{Transport functions for the hypercubic lattice}
\subsection{Conductivity transport function $\Phi_{xx}(\varepsilon)$}
The dispersion relation for an hypercubic lattice in $\text{d}$ dimensions is given by $\varepsilon_k = -2t\sum_{i=1}^{\text{d}}\cos (k_i)$ where we take the lattice constant $a$ as unity. Differentiating $\varepsilon_k$,  Eq.~\eqref{Phi_1} for the conductivity and thermopower transport function becomes
\begin{equation}\label{Phi_1_1}
    \Phi_{xx}(\varepsilon)= (2t)^2\sum_k\sin^2(k_1)\delta (\varepsilon-\varepsilon_k).
\end{equation}
There are d sums over wave vector, one for each direction. Each discrete wave vector sums is assumed normalized so that it becomes $\int dk/(2\pi)$ in the continuum. $\Phi_{xx}(\varepsilon)$  can be evaluated by first computing its Fourier transform~\cite{Uhrig:1996}, that we call $X(w)$.
\begin{equation}\label{Fourier_Phi_1}
\begin{split}
    &X(w) = \int_{-\infty}^{\infty}\Phi_{xx}(\varepsilon)\text{e}^{-iw\varepsilon}d\varepsilon\\
    &= (2t)^2\sum_k\sin^2(k_1)\text{e}^{-iw\varepsilon_k}\\ 
    &= (2t)^2\left[\int_{-\pi}^{\pi}\d{dk}{2\pi}\text{e}^{i2tw\cos (k)}\right]^{\text{d}-1}\int_{-\pi}^{\pi}\d{dk}{2\pi}\sin^2(k)\text{e}^{i2tw\cos (k)}\\
    &= (2t)^2J_0^{\text{d}-1}(2tw)\d{J_1(2tw)}{2tw},
\end{split}
\end{equation}
where $J_0(s)$ and $J_1(s)$ are Bessel functions. These expressions are in the literature~\cite{Jarrell:1993} but the last step, namely obtaining the transport function in arbitrary dimension from the inverse Fourier transform of Eq.~\eqref{Fourier_Phi_1} remained to be taken. 
Using the known relation $\d{dJ_0(s)}{ds} = -J_1(s)$, we obtain $\d{d[J_0(2tw)]^{\text{d}}}{d(2tw)} = -\text{d}[J_0(2tw)]^{\text{d}-1}J_1(2tw)$ so that the inverse transform is
\begin{equation}\label{inv_Fourier_Phi_1}
\begin{split}
    \Phi_{xx}(\varepsilon) &= -\d{1}{2\pi\text{d}}\int_{-\infty}^{\infty}\d{1}{w}\d{d[J_0(2tw)]^{\text{d}}}{dw}\text{e}^{iw\varepsilon}dw\\
    &= -\d{1}{d}\int_{-\infty}^{\infty}F(\varepsilon-z)G(z)dz,
\end{split}
\end{equation}
where we used the convolution theorem and defined $F$ by
\begin{equation}\label{F}
    F(\varepsilon) = \d{1}{2\pi}\int dw\text{e}^{iw\varepsilon}\d{1}{w} = \d{i}{2}sgn(\varepsilon),
\end{equation}
with $sgn$ the sign function, while $G(\varepsilon)$ is defined by
\begin{equation}\label{G}
\begin{split}
    G(\varepsilon) &= -i\varepsilon\d{1}{2\pi}\int_{-\infty}^{\infty}dw\text{e}^{iw\varepsilon}J_0^{\text{d}}(2tw)\\
    &=-i\varepsilon\d{1}{2\pi}\int_{-\infty}^{\infty}dw\text{e}^{iw\varepsilon}\sum_k\text{e}^{-iw\varepsilon_k}\\
    &=-i\varepsilon\sum_k\delta(\varepsilon-\varepsilon_k) = -i\varepsilon N_0(\varepsilon),
\end{split}
\end{equation}
where we used the properties of the Fourier transform of a derivative and defined $N_0(\varepsilon)$ as the non-interacting density of states. Since for the hypercubic lattice with nearest-neighbor hopping only, the density of states is even about $\varepsilon=0$, the final result is then Eq.~\eqref{Phi_1_2}
\begin{equation}
    \Phi_{xx}(\varepsilon) = -\d{1}{\text{d}}\int_{-\infty}^{\varepsilon}zN_0(z)dz.
\end{equation}

\subsection{Stress tensor, $\widetilde{\Phi}_{xx}(\varepsilon)$}
We now turn to the calculation of $\widetilde{\Phi}_{xx}(\varepsilon)$. It is obtained from the following relation between $\Phi_{xx}(\varepsilon)$ and $\widetilde{\Phi}_{xx}(\varepsilon)$:
\begin{equation}\label{derive_Phi_1}
\begin{split}
    \d{d\Phi_{xx}(\varepsilon)}{d\varepsilon} &= \sum_k\left(\d{\pd\varepsilon_k}{\pd k_x}\right)^2\d{\pd\delta (\varepsilon-\varepsilon_k)}{\pd\varepsilon}\\
    &= -\sum_k\left(\d{\pd\varepsilon_k}{\pd k_x}\right)^2\d{\pd\delta (\varepsilon-\varepsilon_k)}{\pd k_x}\left(\d{\pd\varepsilon_k}{\pd k_x}\right)^{-1}\\
    &= -\int\d{dk_{\text{d}}\ldots dk_2}{(2\pi)^{\text{d}}}\int\d{dk_x}{2\pi}\d{\pd\varepsilon_k}{\pd k_x}\d{\pd\delta (\varepsilon-\varepsilon_k)}{\pd k_x}.
\end{split}
\end{equation}
Integrating by part we find $\sum_k\d{\pd^2\varepsilon_k}{\pd k_x^2}\delta(\varepsilon-\varepsilon_k)$ so that with the definition Eq.\eqref{stress_1} of $\widetilde{\Phi}_{xx}(\varepsilon)$ we obtain Eq.~\eqref{Phi_2_1}, namely
\begin{equation}
    \widetilde{\Phi}_{xx}(\varepsilon) = \d{d\Phi_{xx}(\varepsilon)}{d\varepsilon} = -\d{1}{\text{d}}\varepsilon N_0(\varepsilon).
\end{equation}
It is important to note that the relation $\widetilde{\Phi}_{xx}(\varepsilon) = {d\Phi_{xx}(\varepsilon)}/{d\varepsilon}$ is valid in general and not only for the hypercubic lattice.

\subsection{Hall transport function $\Phi_{xy}(\varepsilon)$}
The calculation of $\Phi_{xy} (\varepsilon)$ is more involved but we can use the same kind of approach as for $\Phi_{xx}(\varepsilon)$. Evaluating explicitly the derivatives of $\varepsilon_k$, we may write Eq.~\eqref{Phi_3} as
\begin{equation}\label{Phi_3_1}
\begin{split}
    \Phi_{xy}(\varepsilon)&= (2t)^3\sum_k [\sin^2(k_x)\cos(k_y)\\
    &+ \sin^2(k_y)\cos(k_x)] \delta (\varepsilon-\varepsilon_k)\\
    &=2(2t)^3\sum_k\sin^2(k_x)\cos(k_y)\delta (\varepsilon-\varepsilon_k).
\end{split}
\end{equation}
In Fourier space,
\begin{equation}\label{Fourier_Phi_3_1}
\begin{split}
    Y(w) &= \int_{-\infty}^{\infty}\Phi_{xy}(\varepsilon)\text{e}^{-iw\varepsilon}d\varepsilon\\
    &= 2(2t)^3\sum_k\sin^2(k_1)\cos(k_2)\Pi_{\alpha = 1}^{\text{d}}\text{e}^{i2tw\cos(k_{\alpha})}\\
    &= 2(2t)^3i\left[J_0(2tw)\right]^{\text{d}-2}\d{1}{2tw}\left[J_1(2tw)\right]^2.
\end{split}
\end{equation}
Evaluation of $\d{d^2J_0^\text{d}(2tw)}{dw^2}$ will help us rewrite this expression. Properties of Bessel functions allow us to write
\begin{equation}\label{J0dm2J12}
\begin{split}
    &\left[J_0(2tw)\right]^{\text{d}-2}\left[J_1(2tw)\right]^2 = \d{1}{(2t)^2\text{d}(\text{d}-1)}\d{d^2J_0^\text{d}(2tw)}{dw^2}\\ &+ \d{1}{\text{d}-1}J_0^\text{d}(2tw) - \d{1}{2t(\text{d}-1)}\d{1}{w}\left[J_0(2tw)\right]^{\text{d}-1}J_1(2tw).
\end{split}
\end{equation}
Eq.~\eqref{Fourier_Phi_3_1} can then be rewritten
\begin{equation}\label{Fourier_Phi_3_2}
\begin{split}
    Y(w) &= \d{2i}{\text{d}(\text{d}-1)}\d{1}{w}\d{d^2J_0^\text{d}(2tw)}{dw^2} + \d{2(2t)^2}{\text{d}-1}\d{i}{w}J_0^\text{d}(2tw)\\
    &- \d{2(2t)}{\text{d}-1}\d{i}{w^2}\left[J_0(2tw)\right]^{\text{d}-1}J_1(2tw)\\
    &\equiv Y_1(w) + Y_2(w) + Y_3(w).
\end{split}
\end{equation}
The value of $\Phi_{xy}(\varepsilon)$ is then given by the sum of the inverse Fourier transform of each term in Eq.~\eqref{Fourier_Phi_3_2}. The first term on the right hand side of Eq.~\eqref{Fourier_Phi_3_2} upon inverse transform can, once again, be written as a convolution that is called $\Phi_{xy}^{(1)}(\varepsilon)$. One of the terms in the convolution is given by Eq.~\eqref{F} while the other one coming from the second derivative gives $-\varepsilon^2 N_0(\varepsilon)$. Thus, the convolution can be written as																									
\begin{equation}\label{phi_3_1}
    \Phi_{xy}^{(1)}(\varepsilon) = \d{1}{\text{d}(\text{d}-1)}\left[\int_{-\infty}^{\varepsilon}z^2N_0(z)dz - \int_{\varepsilon}^{\infty}z^2N_0(z)dz\right].
\end{equation}
We can easily show that $\int_{-\infty}^{\infty}z^2N_0(z)dz = \sum_k\varepsilon_k^2 = 2t^2\text{d}$ for the hypercubic lattice. Therefore we obtain
\begin{equation}\label{phi_3_1_2}
    \Phi_{xy}^{(1)}(\varepsilon) = \d{1}{\text{d}(\text{d}-1)}\left[2\int_{-\infty}^{\varepsilon}z^2N_0(z)dz - 2t^2\text{d}\right].
\end{equation}

The second term $\Phi_{xy}^{(2)}(\varepsilon)$ on the right hand side of Eq.~\eqref{Fourier_Phi_3_2} can also be written as a convolution of inverse transforms. The term with $J_0$ gives $N_0(\varepsilon)$ and the other term is again Eq.~\eqref{F}. We thus may write
\begin{equation}\label{phi_3_2}
    \Phi_{xy}^{(2)}(\varepsilon) = -\d{(2t)^2}{(\text{d}-1)}\left[\int_{-\infty}^{\varepsilon}N_0(z)dz - \int_{\varepsilon}^{\infty}N_0(z)dz\right].
\end{equation}
But we know that the integral of the DOS over all frequencies is equal to one $\int_{-\infty}^{\infty}N_0(z)dz = 1$ and thus
\begin{equation}\label{phi_3_2_1}
    \Phi_{xy}^{(2)}(\varepsilon) = -\d{(2t)^2}{(\text{d}-1)}2\int_{-\infty}^{\varepsilon}N_0(z)dz + \d{(2t)^2}{(\text{d}-1)}.
\end{equation}

Finally for $\Phi_{xy}^{(3)}(\varepsilon)$ we find the convolution,
\begin{equation}\label{phi_3_3}
\begin{split}
    \Phi_{xy}^{(3)}(\varepsilon) &= \d{2}{\text{d}(\text{d}-1)}i\d{1}{2\pi}\int_{-\infty}^{\infty}\d{1}{w^2}\d{d\left[ J_0(2tw) \right]^\text{d}}{dw}\text{e}^{iw\varepsilon}\\
    &= \d{2}{\text{d}(\text{d}-1)}iH(\varepsilon)\ast G(\varepsilon),
\end{split}
\end{equation}
where $G(\varepsilon)$ is given by Eq.~\eqref{G} and $H(\varepsilon) = \d{1}{2\pi}\int\d{1}{w^2}\text{e}^{iw\varepsilon}dw = -\d{\varepsilon}{2}sgn(\varepsilon)$. This leads to
\begin{equation}\label{Phi_3_3_1}
\begin{split}
    \Phi_{xy}^{(3)}(\varepsilon) &= -2\d{\varepsilon}{\text{d}(\text{d}-1)}\int_{-\infty}^{\varepsilon}zN_0(z)dz\\ &+ \d{2}{\text{d}(\text{d}-1)}\int_{-\infty}^{\varepsilon}z^2N_0(z)dz - \d{2t^2}{(\text{d}-1)}.
\end{split}
\end{equation}

The final result Eq.~\eqref{phi_3} is obtained by adding $\Phi_{xy}^{(1)}$, $\Phi_{xy}^{(2)}$ and $\Phi_{xy}^{(3)}$,
\begin{equation}
\begin{split}
    \Phi_{xy}(\varepsilon) &= \d{2\varepsilon}{(\text{d}-1)}\Phi_{xx}(\varepsilon)\\ &+ \d{4}{\text{d}(\text{d}-1)}\int_{-\infty}^{\varepsilon}z^2N_0(z)dz\\ &- 2\d{(2t)^2}{(\text{d}-1)}\int_{-\infty}^{\varepsilon}N_0(z)dz.
\end{split}
\end{equation}
We have thus succeeded in writing sums over delta functions in $k$ space as one dimensional integrals over the density of states in energy space.\\
\section{Interacting transport functions}
In this appendix, we give details of the derivation for the longitudinal and transverse interacting transport functions Eqs.~\eqref{tau_tr} and \eqref{tau_tr_H}.

\subsection{Longitudinal conductivities}

Following Ref.~[\onlinecite{Freericks:2006}], Eq.~\eqref{tautr} for $\Phi_{tr}^L (\omega)$
\begin{equation}
\Phi_{tr}^L(\omega) =\frac{1}{\pi^2}\int d\varepsilon \Phi_{xx}(\varepsilon)\left[\text{Im}\left\{\frac{1}{(z-\varepsilon)}\right\}\right]^2,
\end{equation}
where $z \equiv \omega + \mu - \Sigma (\omega)$, can be rewritten with the help of the Green functions Eqs.~\eqref{GLtr} and their derivatives since
\begin{equation}\label{dReG_tr}
\begin{split}
  \text{Re}\left\{ \d{\pd G_{tr}^L(z)}{\pd z} \right\} &= -\int d\varepsilon\d{\Phi_{xx}(\varepsilon)}{|z|^2 - 2\varepsilon\text{Re}\{z\}+\varepsilon^2}\\ &+ 2\int d\varepsilon\d{\Phi_{xx}(\varepsilon)\left(\text{Im}\{z\}\right)^2}{\left(|z|^2 - 2\varepsilon\text{Re}\{z\}+\varepsilon^2\right)^2}
\end{split}
\end{equation}
and
\begin{equation}\label{ImGtr}
  \text{Im}\left\{G_{tr}^L(z)\right\} = \text{Im}\{z^*\}\int d\varepsilon\d{\Phi_{xx}(\varepsilon)}{|z|^2 - 2\varepsilon\text{Re}\{z\}+\varepsilon^2}.\nonumber
\end{equation}
Finally, using integration by part and Eq.~\eqref{Phi_2_1} for  $\frac{d\Phi_{xx}(\varepsilon)}{d \varepsilon}$, one can show that
\begin{equation}\label{dGtrdz}
\d{\pd G_{tr}^L(z)}{\pd z} = \d{1}{\text{d}}\left(1- zG(z)\right)
\end{equation}
and obtain the final result\cite{Freericks:2006} Eq.~\eqref{tau_tr}.


\subsection{Transverse conductivities}
The case of the transverse conductivities is similar to the above. We wish to rewrite the transverse transport function as defined from Eq.~\eqref{tautrH} by
\begin{equation}\label{tautrH_appen}
  \Phi_{tr}^T(\omega) \equiv -\d{1}{\pi^3}\int d\varepsilon\Phi_{xy}(\varepsilon)\left[\text{Im}\left\{\d{1}{z-\varepsilon}\right\}\right]^3.
\end{equation}
We first define a transverse transport Green's function in the form
\begin{equation}\label{GtrH}
  G_{tr}^T(z) = \int d\varepsilon \d{\Phi_{xy}(\varepsilon)}{z-\varepsilon}.
\end{equation}

Here we will need both the first and the second derivative of $G_{tr}^T$. Integrating by parts, computing $\frac{d\Phi_{xy}(\varepsilon)}{d \varepsilon}$ for the hypercubic latice from Eq.~\eqref{phi_3}, using Eq.~\eqref{dGtrdz} and $\Phi_{xx}(\infty)=0$ we find
\begin{equation}\label{deriv_GtrH}
\begin{split}
  \d{\pd G_{tr}^T (z)}{\pd z} &= -\int d\varepsilon\d{\Phi_{xy}(\varepsilon)}{(z-\varepsilon)^2}\\
  &= -\d{2z}{\text{d}(\text{d}-1)}(1-zG(z)) + \d{2}{\text{d}-1}G_{tr}^L(z)\\ &- \d{2(2t)^2}{\text{d}-1}G(z).
\end{split}
\end{equation}
Taking the derivative of that equation with respect to $z$ and using Eq.~\eqref{dGtrdz} the second derivative takes the form
\begin{equation}\label{deriv2_GtrH}
  \d{\pd^2 G_{tr}^T (z)}{\pd z^2}
  = \d{2z}{\text{d}(\text{d}-1)}G(z) + \left[\d{2z^2}{\text{d}(1-\text{d})} -\d{2(2t)^2}{\text{d}-1} \right]\d{\pd G(z)}{\pd z}.
\end{equation}

To see that the interacting transport function can be rewritten with the above derivatives, note that
\begin{equation}\label{Imd2GtrH}
\begin{split}
  \text{Im}\left\{  \d{\pd^2 G_{tr}^T (z)}{\pd z^2}   \right\}&= \text{Im}\left\{ 2\int d\varepsilon\d{\Phi_{xy}(\varepsilon)}{(z-\varepsilon)^3}\right\} \\
  = &8\pi^3\Phi_{tr}^T(z)\\ &+ \d{6}{\text{Im}\{z^*\}}\int d\varepsilon\d{\Phi_{xy}(\varepsilon)\left(\text{Im}\{z^*\}\right)^2}{\left(|z|^2 - 2\varepsilon\text{Re}\{z\}+\varepsilon^2\right)^2}.
\end{split}
\end{equation}
Using the expressions for $\text{Re}\left\{ \d{\pd G_{tr}^T(z)}{\pd z} \right\}$ and $\text{Im}\left\{G_{tr}^T(z)\right\}$ that can be obtained by replacing $\Phi_{xx}$ by $\Phi_{xy}$ in Eqs.~\eqref{dReG_tr} and \eqref{ImGtr}, Eq.~\eqref{Imd2GtrH} takes the form
\begin{equation}\label{Imd2GtrH_1}
\begin{split}
  -8\pi^3\Phi_{tr}^T(z) &= - \text{Im}\left\{  \d{\pd^2 G_{tr}^T (z)}{\pd z^2}   \right\}  - \d{3}{\text{Im}\{z\}}\text{Re}\left\{ \d{\pd G_{tr}^T(z)}{\pd z} \right\}\\ &+ \d{3}{\left(\text{Im}\{z\}\right)^2}\text{Im}\left\{G_{tr}^T(z)\right\}.
\end{split}
\end{equation}

To obtain the desired result Eq.~\eqref{tau_tr_H}, we finally replace the derivatives of $G_{tr}^T(z)$ by the expressions Eqs.~\eqref{deriv_GtrH} and \eqref{deriv2_GtrH} that we obtained above for the hypercubic lattice.


\end{document}